\documentclass[a4paper,11pt]{article}
\pdfoutput=1 % if your are submitting a pdflatex (\emph{i.e.} if you have
\usepackage[utf8]{inputenc}

\usepackage[no-natbib-sort]{jheppub}
\usepackage{enumitem}

\usepackage{youngtab}
\usepackage{hyperref}
\usepackage{fancybox}
\usepackage{float}
\usepackage{amsfonts}
\usepackage{amsmath}
\usepackage{amsbsy}
\usepackage{graphicx}
\usepackage{amssymb}
\usepackage{amsthm}
\usepackage{bm}
\usepackage{comment}
\usepackage{epsfig}
\usepackage{latexsym}
\usepackage{pdflscape}
\usepackage{cancel} % Dirac slash   https://jansoehlke.com/2010/06/strikethrough-in-latex/
\usepackage{color}
\usepackage{cleveref}
\usepackage{bm,nicefrac}


\DeclareSymbolFont{AMSa}{U}{msa}{m}{n}
\DeclareSymbolFont{AMSb}{U}{msb}{m}{n}
\DeclareMathSymbol{\fieldR}{\mathalpha}{AMSb}{"52}

\newcommand{\beq}{\begin{eqnarray}}
\newcommand{\eeq}{\end{eqnarray}}
\newcommand{\bea}{\begin{eqnarray}}
\newcommand{\eea}{\end{eqnarray}}
\newcommand{\be}{\begin{equation}}
\newcommand{\ee}{\end{equation}}
\newcommand{\bq}{\begin{equation}}
\newcommand{\eq}{\end{equation}}

\def\dd{{\rm d}}

\usepackage{color}
    \definecolor{darkgreen}{rgb}{0,0.5,0}
    \definecolor{darkblue}{rgb}{0,0,0.6}
    \definecolor{purple}{rgb}{0.4,.2,0.7}

\newcommand{\fig}[1]{Fig.~\ref{#1}}

\newcommand{\eqn}[1]{(\ref{#1})}

\newcommand{\sac}{\, , \qquad}

\title{Holographic Evolution with Dynamical Boundary Gravity}

\author[a]{Christian Ecker,}
\author[b]{Wilke van der Schee,}
\author[c,d]{David Mateos,}
\author[c]{Jorge Casalderrey-Solana,}

\affiliation[a]{Institut f\"ur Theoretische Physik, Goethe Universit\"at, Max-von-Laue-Str.~1, 60438 Frankfurt am Main, Germany}
\affiliation[b]{Theoretical Physics Department, CERN, CH-1211 Gen\`eve 23, Switzerland}
\affiliation[c]{Departament de F\'\i sica Qu\`antica i Astrof\'\i sica and Institut de Ci\`encies del Cosmos (ICC),\\  Universitat de Barcelona, Mart\'\i\  i Franqu\`es 1, ES-08028, Barcelona, Spain}
\affiliation[e]{Instituci\'o Catalana de Recerca i Estudis Avan\c cats (ICREA), Passeig Llu\'\i s Companys 23, \\ ES-08010, Barcelona, Spain}

\emailAdd{ecker@itp.uni-frankfurt.de}
\emailAdd{wilke.van.der.schee@cern.ch}
\emailAdd{dmateos@fqa.ub.edu}
\emailAdd{jorge.casalderrey@ub.edu}

\begin{abstract}

\end{abstract}

\abstract{Holography has provided valuable insights into the time evolution of strongly coupled gauge theories in a fixed spacetime. However, this framework is insufficient if this spacetime is dynamical. We present a scheme to evolve a four-dimensional, strongly interacting gauge theory coupled to four-dimensional dynamical gravity in the semiclassical regime. As in previous work, we use holography to evolve the quantum gauge theory stress tensor, whereas the four-dimensional metric evolves according to Einstein's equations coupled to the expectation value of the stress tensor. The novelty of our approach is that both the boundary and the bulk spacetimes are constructed dynamically, one time step at a time. We focus on Friedmann-Lemaître-Robertson-Walker geometries and evolve far-from-equilibrium initial states that lead to asymptotically  expanding, flat or collapsing Universes.}

\begin{document} 
\maketitle
\flushbottom

%%%%%%%%%%%%%%%%%%%%%%%%%%%
\section{Introduction}
\label{intro}
%%%%%%%%%%%%%%%%%%%%%%%%%%%%
Holography relates the quantum-mechanical time evolution of a strongly coupled, four-dimensional gauge theory to that of classical gravity in a five-dimensional asymptotically anti de Sitter (AAdS) spacetime.
The power of this correspondence is that it  allows for the use of classical gravity in five dimensions to tackle otherwise intractable problems on the gauge theory side.

The spacetime where the gauge theory is formulated is identified with the boundary of AAdS.
We will refer to its four-dimensional metric as the ``boundary metric'', and to the five-dimensional metric in AAdS as the ``bulk metric''.
In many applications of holography the boundary metric is taken to be non-dynamical.
For example, this metric is flat in the holographic description of the quark-gluon plasma \cite{Casalderrey-Solana:2011dxg,Busza:2018rrf} or in applications to condensed matter systems \cite{Zaanen:2015oix,Hartnoll:2016apf,Nastase:2017cxp}.
Applications with a curved metric include gauge dynamics in black hole backgrounds \cite{Marolf:2013ioa} or in de Sitter (dS) space \cite{Buchel:2002wf,Maldacena:2012xp,Fischler:2013fba,Buchel:2017qwd,Buchel:2017lhu,Buchel:2019pjb,Buchel:2019qcq,Casalderrey-Solana:2020vls}. In all these cases the boundary metric influences, but is unaffected by, the gauge theory dynamics. In other words, the backreaction of the gauge degrees of freedom on the metric is not included.

Despite its successes, this framework is insufficient if the boundary metric is dynamical. This limits potential applications of holography to cosmological defects, phase transitions in the early Universe, neutron star mergers, inflation, pre- or re-heating, cosmological instabilities, etc. In these applications one is interested in the semiclassical-gravity regime. This means that the gauge theory is quantum mechanical but the metric obeys  the classical Einstein equations sourced by the expectation value of the gauge theory stress tensor:
\be
R_{\mu\nu}-\frac{1}{2}R \, g_{\mu\nu} + \Lambda \, g_{\mu\nu} = 
8\pi  G \, \langle  T_{\mu\nu} \rangle \,.
\label{EE}
\ee
All quantities in this equation, including Newton's constant $G$ and a possible cosmological constant $\Lambda$, refer to the four-dimensional boundary theory.
Hereafter we will refer to the gauge theory stress tensor simply as ``the stress tensor''.
Since this is $\mathcal{O}(N^2)$ in the large-$N$ limit, we assume that $G$ is $\mathcal{O}(N^{-2})$ in order to have a finite back-reaction.
In the following we work with $N$-independent quantities defined via the rescalings 
\be
\label{resc}
T_{\mu\nu} \to \left( 2\pi^2 / N^2 \right) T_{\mu\nu} \sac G \to \left( N^2/2\pi^2 \right) G \,.
\ee

\begin{figure}[t]
	\begin{center}
    \includegraphics[width=0.7\textwidth]{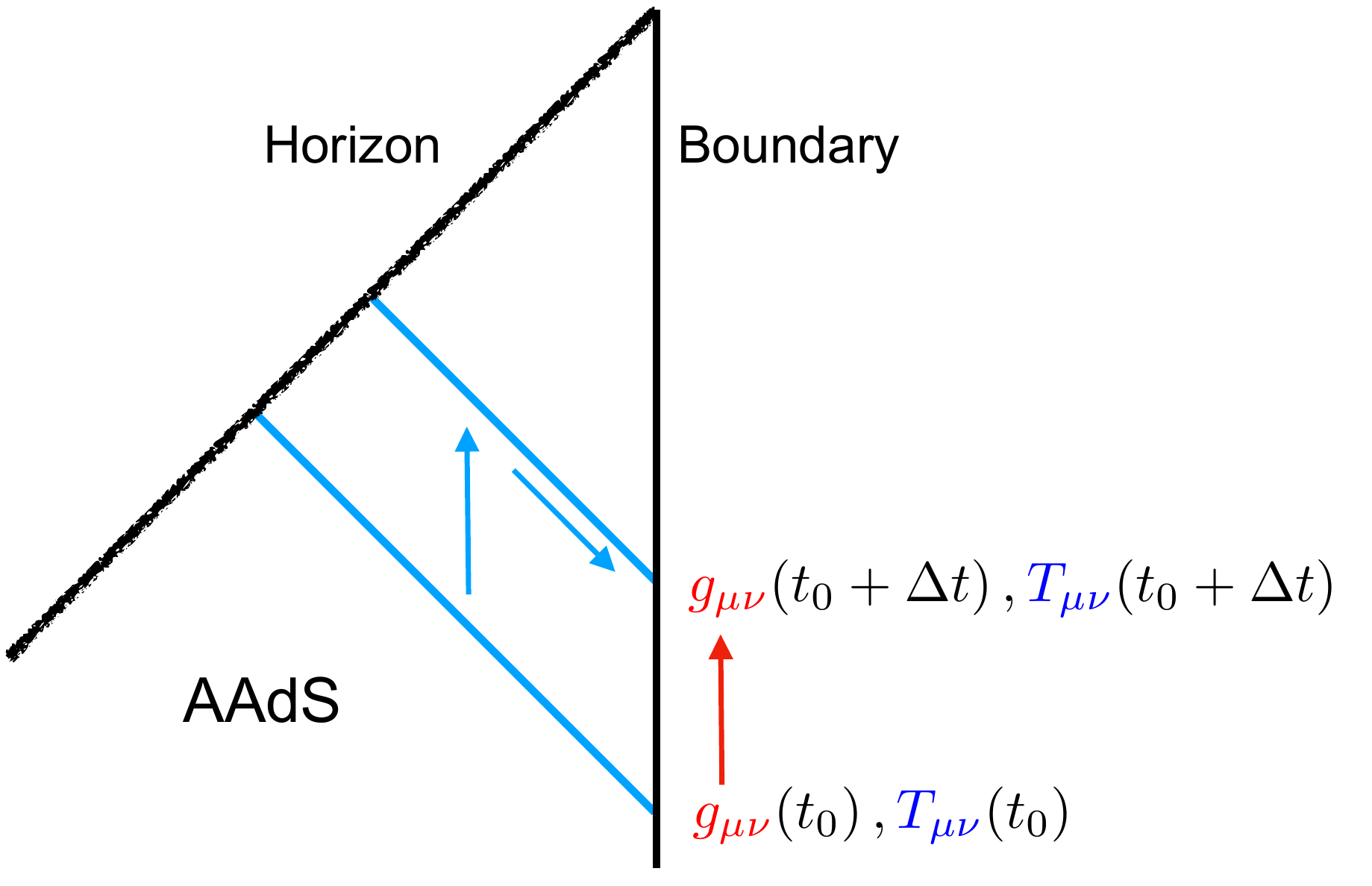}
    \caption{\small Penrose diagram of our evolution scheme.
    The diagonal blue lines are four-dimensional null slices in the bulk.
    Each point on the vertical black line is a three-dimensional spatial slice of the boundary spacetime.
    \label{fig:strategy}}
    	\end{center}
\end{figure}

The key point in the semiclassical regime is to determine the quantum-mechanical evolution of the stress tensor, which must be done self-consistently in the presence of the dynamical metric $g_{\mu\nu}$.
We use holography to determine this evolution (see \fig{fig:strategy}).
The initial state at time $t_0$ is defined by the five-dimensional fields on a bulk null slice, together with the four-dimensional metric on a  boundary  spatial slice.
These two sets of initial data must satisfy non-trivial ``corner'' consistency conditions  that we will analyse below (see \cite{Friedrich:1995vb,Enciso:2014lwa,Carranza:2018wkp,Horowitz:2019dym} 
for related discussions).
For the moment, it suffices to say that the leading term in the near-boundary fall-off of the bulk metric must coincide with the boundary metric, whereas the subleading term in this fall-off determines the expectation value of the stress tensor.
To evolve to a time $t_1=t_0+\Delta t$, we first use equation \eqn{EE} to determine the new boundary metric at $t_1$.
Because AAdS is not globally hyperbolic, this new metric provides necessary boundary conditions that allow us to  evolve the five-dimensional bulk equations to determine the new bulk  fields at $t_1$.
The subleading term of the five-dimensional metric near the boundary then determines the stress tensor at $t_1$. 

The semiclassical regime has been previously extensively considered  in the holographic context. An incomplete list of references includes \cite{Csaki:1999jh,Kehagias:1999vr,Cline:1999ts,Csaki:1999mp,Gubser:1999vj,
Dvali:2000rv,Karch:2000ct,Kiritsis:2005bm,Apostolopoulos:2008ru,Compere:2008us,Erdmenger:2011sy,Dong:2011uf,Banerjee:2012dw,Fischetti:2014hxa,Buchel:2017pto,Chesler:2020exl,Ghosh:2020qsx,Emparan:2020znc}. The main novelty of our approach with respect to previous work is that both the boundary and the bulk spacetimes are constructed dynamically, one time step at a time. Further differences include the fact that we do not introduce an ultraviolet cut-off in the gauge theory or branes in the bulk \cite{Csaki:1999jh,Kehagias:1999vr,Cline:1999ts,Csaki:1999mp,Gubser:1999vj, Dvali:2000rv,Karch:2000ct,Kiritsis:2005bm,Dong:2011uf,Emparan:2020znc} but work directly with dynamical gravity at the boundary; 
%As a consequence, we have complete freedom to choose the dynamical equations obeyed by the boundary metric. 
we do not assume a perfect-fluid form for the stress tensor \cite{Fischetti:2014hxa} or consider a derivative expansion \cite{Buchel:2016cbj} but allow for arbitrarily-far-from-equilibrium dynamics; and we do not make use of predetermined bulk solutions \cite{Apostolopoulos:2008ru,Erdmenger:2011sy,Banerjee:2012dw} or restrict ourselves to constant-curvature boundary metrics \cite{Ghosh:2020qsx}.

For simplicity, in this paper we will focus on homogenous and isotropic gauge-theory states, namely on Friedmann-Lemaître-Robertson-Walker geometries. However, we expect that our approach can be extended to more general cases.  Our full code is publicly available at \href{http://wilkevanderschee.nl/public-codes}{http://wilkevanderschee.nl/public-codes}.

%Our strategy differs from previous work in essential ways.  We do not introduce an ultraviolet cut-off in the gauge theory or branes in the bulk \cite{Csaki:1999jh,Kehagias:1999vr,Cline:1999ts,Csaki:1999mp,Gubser:1999vj, Dvali:2000rv,Karch:2000ct,Kiritsis:2005bm,Dong:2011uf,Emparan:2020znc} but work directly with dynamical gravity at the boundary. As a consequence, we have complete freedom to choose the dynamical equations obeyed by the boundary metric. We do not assume a perfect-fluid form for the stress tensor \cite{Fischetti:2014hxa} or consider a derivative expansion \cite{Buchel:2016cbj} but allow for arbitrarily-far-from-equilibrium dynamics. We do not make use of predetermined bulk solutions \cite{Apostolopoulos:2008ru,Erdmenger:2011sy,Banerjee:2012dw} or restrict ourselves to constant-curvature boundary metrics \cite{Ghosh:2020qsx}, but instead construct both the bulk and the boundary geometries dynamically, one time step at a time.

%%%%%%%%%%%%%%%%%%%%%%%%%%%
\section{Model, scheme and evolution}
\label{model}
%%%%%%%%%%%%%%%%%%%%%%%%%%%%
\subsection{Model}
We use the same model as in Ref.~\cite{Casalderrey-Solana:2020vls}, to which we refer the reader for additional details. The four-dimensional gauge theory is a large-$N$, strongly coupled, non-conformal theory with a mass scale $M$. We will measure all the dimensionful gauge-theory quantities in units of $M$. The five-dimensional bulk theory consists of gravity coupled to a scalar field $\phi$, with action
\begin{equation}\label{action}
S=\frac{2}{8\pi G_5}\int_{\mathcal{M}} \dd^5x\sqrt{- \mathfrak{g}}\left(
\frac{1}{4}\mathcal{R}-\frac{1}{2}(\partial\phi)^2-V(\phi)\right)+\frac{1}{8\pi G_5}\int_{\partial\mathcal{M}}\dd^4x\sqrt{-\gamma}K+S_{ct}\,.
\end{equation}
Here $G_5$ is the five-dimensional Newton's constant, $\mathcal{R}$ is the Ricci scalar associated to the five-dimensional bulk metric $\mathfrak{g}$ on $\mathcal{M}$, $\gamma_{\mu\nu}$ is the metric induced  on a four-dimensional slice near the boundary  $\partial \mathcal{M}$, and 
\be
K=\gamma^{\mu\nu}K_{\mu\nu}=\gamma^{\mu\nu}\nabla_\mu n_\nu
\ee
is the trace of the extrinsic curvature $K_{\mu\nu}$ associated to this slice.
The second term on the right-hand side of \eqn{action} is the familiar Gibbons--Hawking term. The third term in \eqn{action} will be described shortly.
The potential $V(\phi)$ encodes the properties of the dual gauge theory. As in  \cite{Casalderrey-Solana:2020vls}, we choose 
\be
\label{eq:defW}
V(\phi)= -\frac{4}{3} W\left(\phi \right)^2 + \frac{1}{2} W'\left(\phi\right)^2 \,,
\ee 
where the superpotential is given by
\be
\label{eq:defWs}
L\, W\left(\phi \right)=-\frac{3}{2} - \frac{\phi^2}{2} + \frac{\phi^4}{4 \phi^2_M} \,.
\ee
$L$ is a length scale.
The dimensionless constant $\phi_M$ is a free parameter that controls the degree of non-conformality of the model, for example the maximum value of the bulk viscosity.
For concreteness, in this paper we will choose
\be
\phi_M=2 \,.
\ee
Both $V(\phi)$ and $W(\phi)$ have a maximum at $\phi=0$ and a minimum at $\phi=\phi_M$.
Each of these extrema yields an AdS solution of the equations of motion with constant $\phi$ and radius $L^2=-3/V(\phi)$.
In the gauge theory each of these solutions is dual to a fixed point of the renormalisation group with a number of degrees of freedom $N^2$ proportional to $L^3/G_5$.
In top-down models this relation is known precisely.
For example, in the case in which the gauge theory is $\mathcal{N}=4$ SYM with $N$ colours we would have 
\be
\label{defN}
\frac{L^3}{8\pi G_5}= \frac{N^2}{4\pi^2} \,.
\ee
In our bottom-up model we will take this as a definition of the number of degrees of freedom in the gauge theory, $N$, at each fixed point.

\subsection{Scheme}
Near the AdS boundary the metric can be written in the so-called Fefferman--Graham (FG) gauge 
\begin{equation}\label{metricFG}
ds^2= L^2 \, \frac{\dd\rho^2}{4\rho^2}+\gamma_{\mu\nu}(\rho,x)\dd x^\mu \dd x^\nu \,. 
\end{equation}
The boundary is located at $\rho=0$ and is parametrised by the coordinates $x^\mu$ with \mbox{$\mu=0, \ldots, 3$}. Near the boundary the metric and the scalar behave as 
\be
\label{seriesVac}
\gamma_{\mu\nu}(z,x) \sim \frac{g_{\mu\nu}(x)}{\rho} \sac \phi \sim M \rho^{1/2} \,,
\ee
where $g_{\mu\nu}(x)$ is the boundary metric and $M$ is the gauge theory intrinsic scale. 
Substituting this in the first term of the action \eqn{action} we see that it suffers from large-volume divergences.  These divergences can be regularised and renormalised by a procedure called holographic renormalisation (see e.g.~\cite{deHaro:2000vlm,Bianchi:2001de,Bianchi:2001kw}), which makes the action finite and the variational principle well-defined. 
This procedure is implemented  by including in \eqn{action} the counterterm action 
\bea
\label{eq:Sct}
 S_{\mathrm{ct}} = \frac{L}{8\pi G_5} \int_{\partial \mathcal{M}} \dd^4x\sqrt{-\gamma}&& \Bigg[  \left(
-\frac{1}{8}R-\frac{3}{2}-\frac{1}{2}\phi^2 \right) 
+ \frac{1}{2} \left( \log \rho\right)  \mathcal{A}+ \\
&& + L^2\Big(  \,\alpha \mathcal{A} + \beta\phi^4  + \varepsilon \,  \phi^2 R + 
\xi_1  R^2   + \xi_2 \,\nabla^2 R + 
\xi_3 \nabla_\mu \nabla_\nu R^{\mu\nu} \Big) \Bigg] ,\qquad\nonumber
\eea
where $\alpha, \beta, \varepsilon, \xi_i$ are real constants and the factors of $L$ are necessary for dimensional reasons. This action is integrated on a timelike, constant-$\rho$ hypersurface near the boundary with induced metric 
$\gamma_{\mu\nu}$. In this and in subsequent equations all metric-dependent terms such as the Ricci scalar $R$, the covariant derivative $\nabla$, 
etc.~are those associated to $\gamma$. The second term of \eqn{action} is also understood to be evaluated on this slice, the first term  of \eqn{action} is understood to be evaluated by integrating down to this slice, and the limit $\rho \to 0$ is understood to be taken at the end of the calculation. 

In \eqref{eq:Sct},  $\mathcal{A}(\gamma_{\mu\nu}, \phi)$ is the so-called conformal anomaly, which in our case is given by 
\be
\mathcal{A}=\mathcal{A}_g + \mathcal{A}_\phi \,,
\ee
where
\be
\label{eq:Ag}
\mathcal{A}_g = \frac{1}{16}(R^{\mu\nu}R_{\mu\nu}-\frac{1}{3}R^2)
\ee
is the holographic gravitational conformal anomaly and 
\be
\label{eq:Aphi}
\mathcal{A}_{\phi}=-\frac{\phi^2}{12}R 
\ee
is the conformal anomaly due to matter.
In these and in subsequent equations all the terms are functionals of the metric $\gamma_{\mu\nu}$ and of the scalar field $\phi$ induced on the $\rho$-hypersurface. However, making use of the near-boundary behaviour  \eqref{seriesVac} we see that the product with the determinant of the induced metric yields a finite contribution in the limit in which the cut-off is removed, since  
\be
\lim_{\rho\to 0} \, \sqrt{-\gamma}\,  
\mathcal{A} \left( \gamma_{\mu\nu}, \phi \right) \,=\, 
\lim_{\rho\to 0} \,\Big[ \rho^{-4} \,  \sqrt{-g} \Big] \Big[ \rho^4  \mathcal{A} 
\left( g_{\mu\nu}, M \right) \Big] \,=\, 
\sqrt{-g} \, \mathcal{A} \left( g_{\mu\nu}, M\right) \,.
\ee 
For this reason we will often think of the anomaly, as well as of other curvature invariants, as functionals of $M$ and the boundary metric $g_{\mu\nu}$. 
 
The fact that $\sqrt{-\gamma}  \mathcal{A}$ yields a finite result has two consequences.
First, it means that the logarithmic term in \eqn{eq:Sct} cancels a purely logarithmic divergence from the bulk action.
The requirement that this cancellation takes place fixes uniquely the form of the anomaly, including the values of all the numerical coefficients in \eqref{eq:Ag} and \eqref{eq:Aphi}.
The presence of this logarithmic term on the gravity side breaks diffeomorphism invariance and is dual to the presence of the conformal anomaly in the dual gauge theory. 

The second consequence is that the anomaly itself, without the $\log$, can be added to the counterterm action with an arbitrary coefficient, which we named $\alpha$ in \eqn{eq:Sct}. It is important to note that not just the anomaly but any local, finite term that is invariant under the symmetries of the theory can be added to the counterterm action with an arbitrary coefficient. The freedom to add these terms  with arbitrary coefficients  is part of the general freedom in the choice of renormalisation scheme. These terms can be constructed out of non-negative powers of the scalar field and of curvature invariants of the induced metric $\gamma_{\mu\nu}$ in such a way that their overall mass dimension is four.\footnote{Derivatives of the scalar field should also be included in situations with non-constant $M$.}  The  second line of \eqn{eq:Sct} is the most general linear combination of terms of this type, except for the Kretschmann scalar $R_{\mu\nu\tau\psi}R^{\mu\nu\tau\psi}$. We have not included the latter because, in four dimensions, the integral
\be
\frac{1}{8\pi^2} \int \sqrt{-\gamma} \left(R^2 - 4 R_{\mu\nu}R^{\mu\nu}+ 
R_{\mu\nu\tau\psi}R^{\mu\nu\tau\psi} \right) =  \chi \,,
\ee
with $\chi$ the Euler character, is a topological invariant. A pedagogical discussion of the independent curvature invariants in arbitrary dimension can be found in \url{http://kias.dyndns.org/crg/invariants.html}.

The coefficients $\alpha$ and $\beta$ play special  roles.  In the first case, this is because $\alpha$ can be shifted by a scale transformation, which is implemented via the following rescaling of the  coordinates 
\begin{equation}
\label{rescresc}
x^\mu \to \lambda x^\mu \,, \qquad \rho \to \lambda^2 \rho \,,
\end{equation}
where $\lambda$ is a positive real number.  It is easy to see that the effect of this transformation is to shift the counterterm action by a term of the form $(\log \lambda) \mathcal{A}$, which in turn can be absorbed through the redefinition $\alpha \to \alpha + \log \lambda$. The freedom to rescale $\rho$, or equivalently to shift $\alpha$, is thus the freedom to choose a renormalisation scale. We may therefore think of $\alpha$ as related to the renormalization group scale or subtraction point $\mu$ through
\be
\label{rg}
\alpha \sim \log \mu \,.
\ee
In the second case the reason is that the counterterm associated to $\beta$ is the only one that does not vanish for a flat boundary metric. In particular, the value 
\be
\label{special}
\beta=\frac{1}{4\phi_M^2}=\frac{1}{16} 
\ee
is special because in this case the $\beta \phi^4$ term combines with the second and the third summands in the first term of \eqn{eq:Sct} to give precisely the superpotential \eqn{eq:defWs}. This means that, if the theory \eqn{action} is the bosonic truncation of a supersymmetric theory with superpotential $W$, then this choice of $\beta$ leads to a supersymmetric renormalisation scheme and, as a consequence, the full boundary stress tensor vanishes identically if the boundary metric is flat. We will come back to these points below.

The finite counterterms give contributions to the gauge theory stress tensor, which therefore we can write as 
\be
T_{\mu\nu}=T_{\mu\nu}^{(0)} + \alpha \left( T_{\mu\nu}^{(g)} + T_{\mu\nu}^{(\phi)} \right) + \beta \, T_{\mu\nu}^{(\beta)} +\varepsilon \, T_{\mu\nu}^{(\varepsilon)} 
+ \xi_i \, T_{\mu\nu}^{(i)}  \,,
\ee
where $T_{\mu\nu}^{(0)}$ denotes the stress tensor in the scheme 
$\alpha=\beta=\varepsilon=\xi_i=0$. This means that, in the absence of dynamical gravity at the boundary, the boundary stress tensor is ambiguous to the extent that the coefficients of the finite counterterms are arbitrary. However, in the presence of dynamical boundary gravity, these coefficients simply renormalize the gravitational couplings and the ambiguity is replaced by the physical specification of the renormalized couplings \cite{Birrell:1982ix}. To see this,  
imagine first setting all coefficients to zero except for $\beta$ and consider  the contribution to the full stress tensor of $T_{\mu\nu}^{(\beta)}$, which takes the form
\be
\label{Tbeta}
T_{\mu\nu}^{(\beta)} =  \frac{L^3}{8\pi G_5} \,M^4\, g_{\mu\nu} 
= \frac{N^2}{4\pi^2}\, M^4\, g_{\mu\nu} \,,
\ee
where we have made use of \eqn{defN}. Moving this term to the left-hand side of \eqn{EE} we can write Einstein's equations in the form  
\be
\label{EE2}
\frac{1}{8\pi G} \left( R_{\mu\nu} - \frac{1}{2} \, R g_{\mu\nu}\right) + 
\frac{\Lambda_{\rm{ren}}}{8\pi G} \, g_{\mu\nu} = T_{\mu\nu}^{(0)} \,,
\ee
with 
\be
\label{lambdaren}
\frac{\Lambda_{\rm{ren}}}{8\pi G}=\frac{\Lambda}{8\pi G}-
\frac{\beta M^4}{4\pi^2}  \,.
\ee
Note that in this equation the $N^2$-factor  coming from \eqn{Tbeta} has cancelled out with the $N^2$-factor coming from \eqn{resc} in such a way that \eqn{lambdaren}  is $N$-independent.  We see that the effect of the $\beta$-counterterm is simply to renormalize the bare cosmological constant\footnote{More precisely, the combination $\Lambda/8\pi G$.} $\Lambda$ in Einstein's equations. In other words, $\Lambda$ and $\beta$ are not separately meaningful, only the combination  $\Lambda_{\rm{ren}}$ is. It is therefore convenient to choose $\beta$ as in \eqn{special}, since in this case flat space is a solution of \eqn{EE} with $\Lambda=0$.  

Consider now adding the contribution of the $\alpha$-term. Also, assume for the moment that the boundary metric is of FRWL type, which will be our focus below. On this subclass of boundary states the stress tensor associated to  $\mathcal{A}_g$ vanishes, namely $T_{\mu\nu}^{(g)}=0$,  and the stress tensor of $\mathcal{A}_\phi$ combines with that of the $\varepsilon$-counterterm to give  
\be
\frac{L^3}{8\pi G_5} \Big[ \alpha \left( T_{\mu\nu}^g + T_{\mu\nu}^\phi \right) +\varepsilon \, T_{\mu\nu}^{(1)}\Big] = 
%\alpha\, T_{\mu\nu}^\phi +\delta\, T_{\mu\nu}^\delta = 
- \frac{N^2}{4\pi^2} \frac{M^2}{6} 
\left( \alpha + 2\varepsilon \right) \left(R_{\mu\nu} - \frac{1}{2} \, R g_{\mu\nu} \right)
\,,
\ee
where we have made use of \eqn{defN}. Moving this term to the left-hand side of \eqn{EE} we get 
\be
\label{EE2}
\frac{1}{8\pi G_{\rm{ren}}} \left( R_{\mu\nu} - \frac{1}{2} \, R g_{\mu\nu}\right) + \frac{\Lambda_{\rm{ren}}}{8\pi G_{\rm{ren}}} \, g_{\mu\nu} = T_{\mu\nu}^{(0)} \,,
\ee
with 
\be
\label{gren}
\frac{1}{8\pi G_{\rm{ren}}} = \frac{1}{8\pi G} +  
\frac{N^2}{4\pi^2} \frac{M^2}{6} 
\left( \alpha + 2\varepsilon \right) \,. \vspace{2mm}
\ee
As in \eqn{lambdaren}, the $N^2$-factors cancel out in \eqn{gren}. We see that the effect of the $\alpha$- and $\varepsilon$-terms is simply to renormalize the bare, four-dimensional  Newton's constant  $G$. In other words, $G$, $\alpha$ and $\varepsilon$ are not separately meaningful, only the combination  $G_{\rm{ren}}$ is. Since the coefficient $\alpha$ is associated to renormalization group transformations through \eqn{rg},  Eq.~\eqn{gren} can be seen as the renormalization group equation for the running of Newton's constant. For convenience we will work in the scheme $\alpha=\varepsilon=0$.
%since in this scheme the constant $G$  in \eqn{EE} is directly the renormalized Newton's constant in the boundary theory. 

Consider now what happens if the boundary metric is not FRWL or if we consider the rest of the finite counterterms with coefficients $\zeta_i$. These terms are of order higher than two in derivatives.  If we were considering the most general semiclassical gravitational theory, as we would do in the full effective theory, then these contributions would simply renormalize the bare values of the coefficients of higher-curvature terms that were omitted in \eqn{EE}. As above, the only  meaningful quantities would be the renormalized couplings measured at the physical energy scale of interest $\mu$.  The dynamical regime that we wish to study is that of classical gravity coupled to quantum matter. This means that $\mu\ll M_p$, with $M_p$ the Planck mass, since otherwise we need to include quantum gravity effects. This is the regime of interest, for example, if we want to understand how the QCD transition took place as the Universe expanded and cooled. In this regime we expect that the renormalized higher-derivative couplings will be suppressed by factors of $\mu/M_p$. Therefore it is a consistent approximation to work with a truncated effective theory in which we set these couplings to zero. Since only renormalized couplings matter we may therefore declare that, in this approximation, all the $\xi_i$ vanish and the left-hand side of \eqn{EE} contains only two-derivative terms.

\subsection{Evolution}
Having fixed the renormalization scheme, we can now discuss the dynamics. For simplicity, we focus on homogeneous and isotropic states in the four-dimensional theory, namely on Friedmann-Lemaître-Robertson-Walker cosmologies.
As a consequence, the boundary metric is completely determined by a scale factor $a(t)$, and the only non-zero components of the stress tensor are the energy density $\mathcal{E}(t)$ and the pressure $\mathcal{P}(t)$.
Under these conditions, \eqn{EE} reduces to the Friedmann equation
\be
\label{F1}
\left( \frac{\dot a}{a} \right)^2 \equiv H^2 = \frac{1}{3}\, \Lambda 
+ \frac{8\pi G}{3}\, \mathcal{E}
\ee
and the continuity equation 
\be
\label{Fcont}
\dot{\mathcal{E}} = - 3H \left(\mathcal{E}+\mathcal{P} \right) \,,
\ee
with $H=\dot a/a$ the Hubble rate.

Below we will extract the stress tensor from the near-boundary fall-off of the five-dimensional fields. For illustration, consider the bulk scalar field. In an appropriate null holographic coordinate $r$ with the boundary at $r\to \infty$ we have \cite{Casalderrey-Solana:2020vls}
\be
\label{expansion}
\phi = \frac{M}{r} + \frac{\phi_2(t)}{r^3}+  \frac{1}{r} \sum_{n \geq 3} \frac{\phi_{n} (t)}{r^{n}} + \frac{1}{r} \sum_{n \geq 2} \frac{\psi_n (t) \log r}{r^{n}} + \cdots
\ee
The logarithmic terms are specific to odd-dimensional bulk spacetimes \cite{AST_1985__S131__95_0}.
The near-boundary analysis only leaves undetermined $\phi_2(t)$. 
The remaining coefficients $\phi_{n\geq 3}$ and $\psi_{n\geq 2}$ are given in terms of $\phi_2(t)$, $a(t)$ and their derivatives by expressions of the form
\begin{subequations}
\label{coef}
\begin{align}
\label{coef1}
& \phi_n \left( M, a, \dot a, \ldots, a^{(n)}, \phi_2, \dot \phi_2, \ldots, \phi_2^{(n-2)}\right), \\
\label{coef2}
&\psi_n \left( M, a, \dot a, \ldots, a^{(n)} \right). \,\,\,\,\,\,\,\,
\end{align}
\end{subequations}

There is a similar expression for the fall-off of the five-dimensional bulk metric with one undetermined coefficient $a_4(t)$.
From the bulk viewpoint, the function $\phi(r,t_0)$ and the coefficient $a_4(t_0)$ at an initial time $t_0$ are free data. 
Moreover, if this data and the scale factor $a(t_0)$ are known, then integration of the constraints coming from the Einstein-scalar equations in the bulk determines the entire five-dimensional metric on the initial time slice at $t=t_0$.
\begin{figure*}[h!!!]
	\begin{center}
    \includegraphics[height=0.43\textwidth]{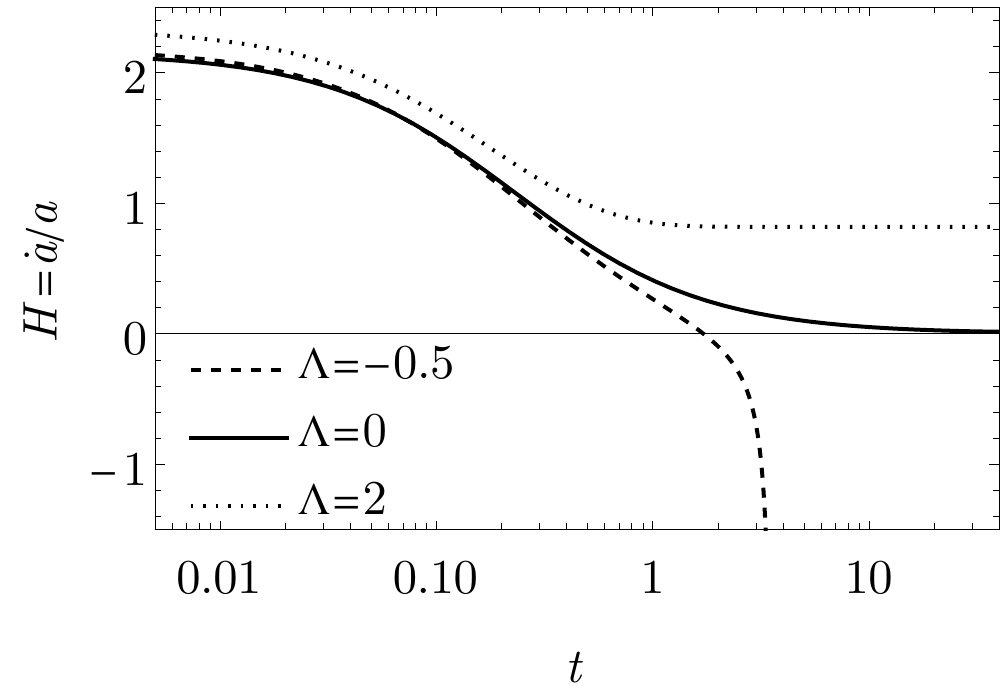} \,\,\,\,\,\, \\
    \includegraphics[height=0.43\textwidth]{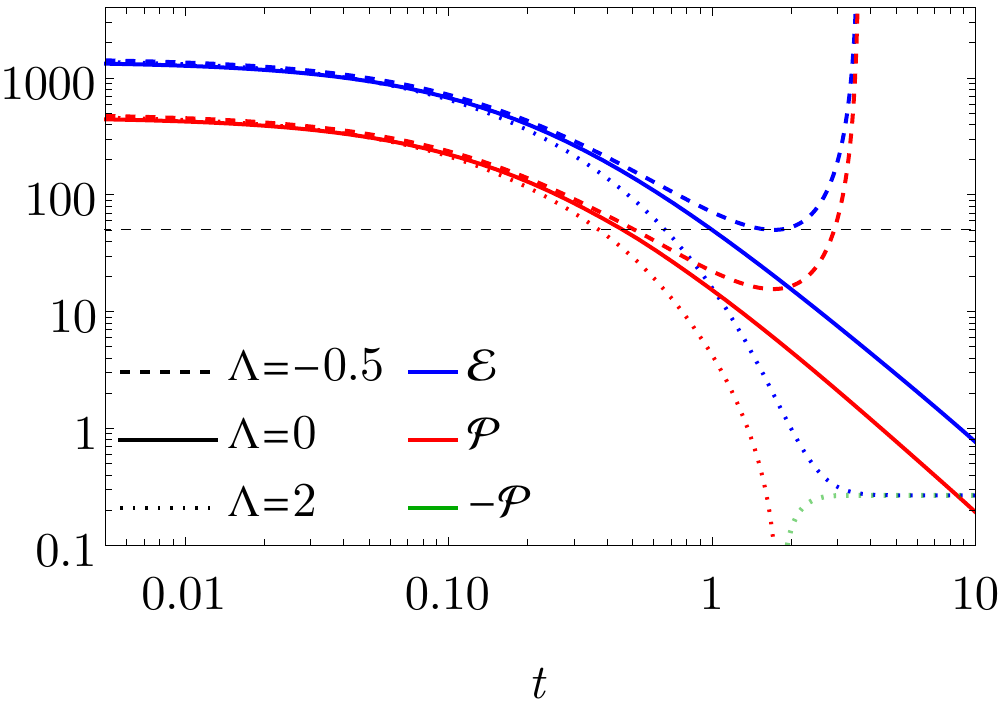} \\
     \includegraphics[height=0.45\textwidth]{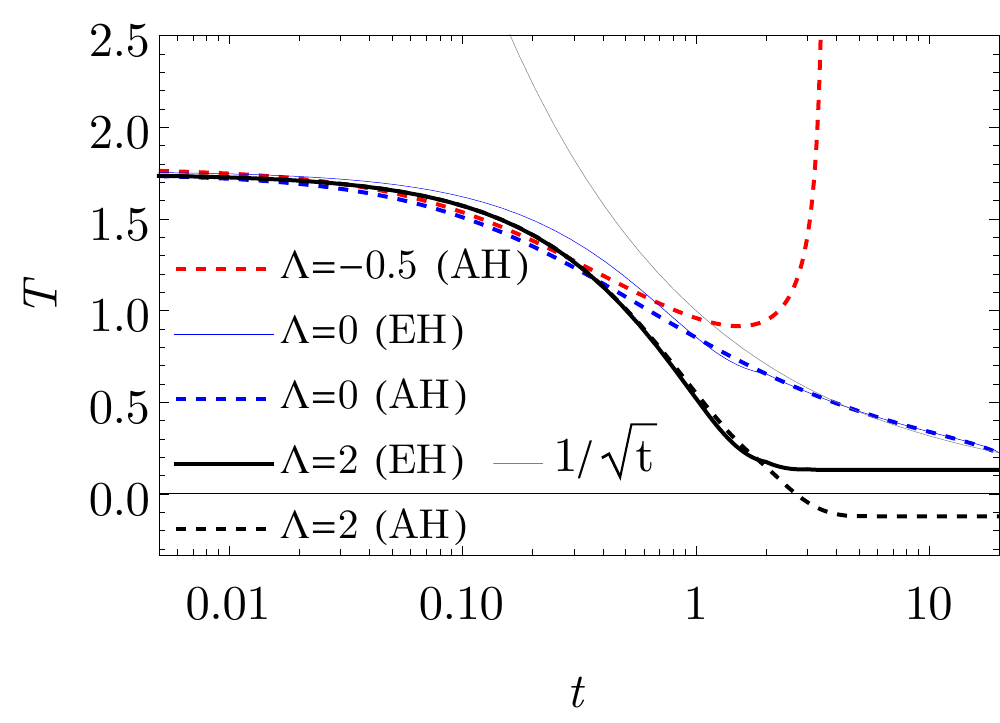} \,\,\,\,\,\,
    \caption{\small Evolution of the Hubble rate (top), of the energy density and pressure (middle), and of the effective temperature (bottom), for $G=1/2500$ and three different values of $\Lambda$. 
    \label{fig:Hoft}}
    	\end{center}
\end{figure*}

\begin{figure}[htb]
	\begin{center}
    \includegraphics[height=0.45\textwidth]{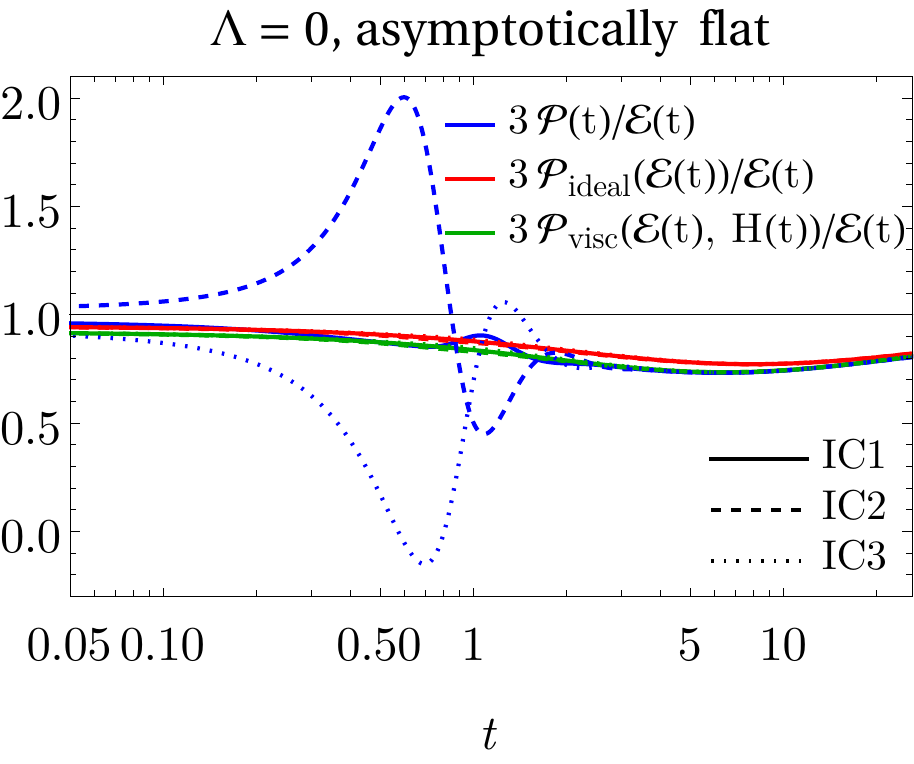}
    \includegraphics[height=0.45\textwidth]{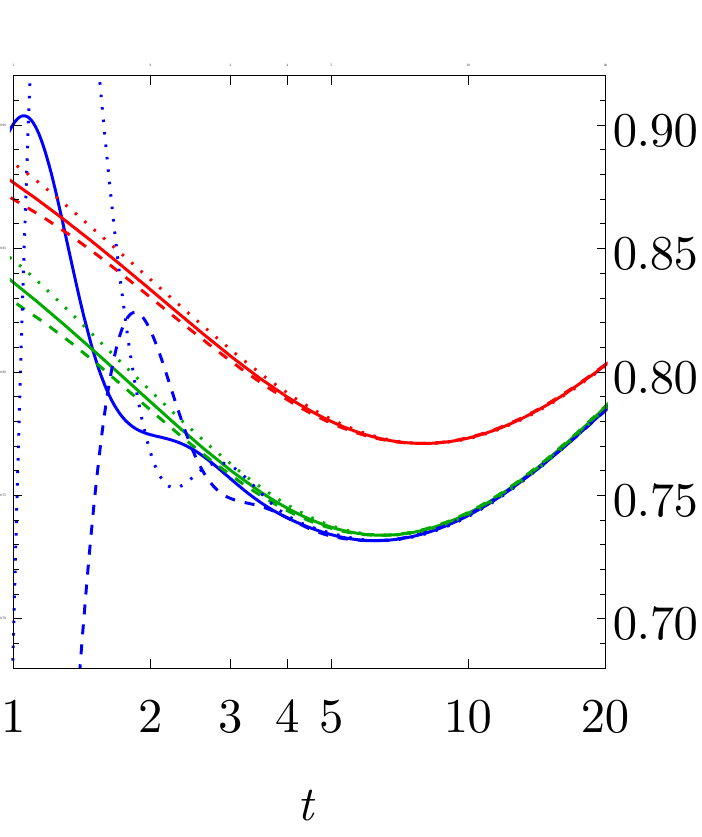}
    \caption{\small 
   Comparison between the holographic result for the pressure/energy ratio (blue) and the ideal (red) and viscous (green) hydrodynamic approximations, for $G=1/750$, $\Lambda=0$ and three different initial conditions.
   \label{fig:pumukel}} 
       	\end{center}
\end{figure}

Eqs.~\eqn{coef}, which arise from the bulk equations of motion, constitute a set of constraints that relate the bulk initial condition $\phi(r,t_0)$ and the derivatives of the boundary scale factor.
These corner conditions imply that the initial data on the bulk slice and the boundary conditions on that slice cannot be specified independently.
For a non-dynamical boundary metric, as is in e.g.~\cite{Chesler:2008hg}, $a(t)$ can be prescribed arbitrarily and these bulk constraints can be used to determine the $\psi_n(t)$ coefficients.  
In contrast, in the case of dynamical boundary gravity, it is highly non-trivial that these bulk constraints can be made compatible with those coming from the boundary Einstein equations \eqn{EE}.
The latter arise as follows. The stress tensor depends on the undetermined coefficients and on the scale factor via expressions of the form \cite{Casalderrey-Solana:2020vls}
\be
\label{EP}
\mathcal{E} \left( a_4, \phi_2, a, \dot a, \ddot a \right) \sac 
\mathcal{P} \left( a_4, \phi_2, a, \dot a, \ddot a \right) \,.
\ee
These, together with \eqn{F1}, \eqn{Fcont} and \eqn{coef1}, can be shown to determine all the derivatives of the scale factor at  $t_0$ in terms of $a(t_0)$, $a_4(t_0)$ and $\phi_n(t_0)$.
Through \eqn{coef2}, this  fixes all the logarithmic terms in $\phi(r,t_0)$.
This is particularly important for our scheme because numerically we work with ``subtracted'' variables that differ from the original ones by a number of logarithmic (and some non-logarithmic) terms.
Specifically, our evolution scheme is as follows.
At $t_0$ we specify $a(t_0)$, $a_4(t_0)$, $\phi_2(t_0)$ and the subtracted version of $\phi(r,t_0)$. 
We then use the procedure above to find $a^{(n)}(t_0)$ up to $n=4$.
These determine all the necessary logarithmic terms.
Next we integrate the Einstein-scalar constraints and find all the bulk data on the initial time slice.
By construction this is consistent with the corner conditions up to the desired order.
Finally, we use the bulk and the boundary evolution equations to obtain $a$, $a_4$, $\phi_2$ and the subtracted version of $\phi$ at $t_0+\Delta t$.

%%%%%%%%%%%%%%%%%%%%%%%%%%%
\section{Results}
\label{results}
%%%%%%%%%%%%%%%%%%%%%%%%%%%%
We perform evolutions for three different values of the cosmological constant $\Lambda=\{-0.5, 0, 2\}$. 
As initial data at $t_0=0$ we use $a(0)=1$ and a radial profile $\phi(r,t_0)$ that corresponds to a thermal equilibrium state in flat space. 
In all cases we choose $a_4(0)=-2000$, except in Fig.~\ref{fig:pumukel}, for which $a_4(0)=-100$.

Fig.~\ref{fig:Hoft}(top) shows the evolution of the Hubble rate.
Negative $\Lambda$ leads to a ``Big Crunch'' where the Hubble rate evolves towards minus infinity and the spacetime collapses.
For $\Lambda=0$ the Hubble rate decays to zero and the spacetime approaches Minkowski space.
Positive $\Lambda$ leads to an exponentially expanding dS Universe.

Fig.~\ref{fig:Hoft}(middle) shows $\mathcal{E}$ and $\mathcal{P}$.
For  $\Lambda<0$ the energy density reaches a minimum, after which it diverges as the Big Crunch is approached.
For  \mbox{$\Lambda=0$}, $\mathcal{E}$ and $\mathcal{P}$ decrease in a power-law fashion that is well described by hydrodynamics (see below).
For $\Lambda>0$ the Universe approaches dS with a small Casimir contribution from the non-conformal matter, \mbox{$\mathcal{E}_{\rm dS} = - \mathcal{P}_{\rm dS} \approx 0.2667$}. \footnote{This value is consistent with \cite{Casalderrey-Solana:2020vls} after taking into account a typo in Sec 4.3 of \cite{Casalderrey-Solana:2020vls}, where we wrote  that we chose $\alpha=0$ while the actual value was $\alpha=3/4$.}

In Fig.~\ref{fig:Hoft}(bottom) we show the temperature of the gauge theory state, $T=\kappa/2\pi$, computed from the surface gravity, $\kappa$, of the event (EH) and of the apparent (AH) horizons of the bulk geometry.
For $\Lambda<0$ the AH reaches the boundary of AAdS at a finite boundary proper time. The boundary itself collapses at this point, and $T_{AH}$ diverges.
We do not show $T_{EH}$ because the definition of the EH is unclear in this case.
For  $\Lambda>0$ the temperatures at late times approach $T_{EH}=-T_{AH}=H/2\pi$, in agreement with \cite{Buchel:2017pto, Casalderrey-Solana:2020vls}.
For $\Lambda=0$ the horizon falls deep into the bulk and at late times 
$H\propto t^{-1}$ and $T\propto t^{-1/2}$, as expected.
In addition, Eqn.~(\ref{F1}) implies $\mathcal{E}/H^4 \sim t^2/G \gg 1$, meaning that the dynamics is dominated by the energy density.
As a consequence, the late-time boundary state approaches a thermal state in Minkowski space and the bulk EH and AH become indistinguishable.

Holography can evolve strongly-coupled, far-from-equilibrium, quantum matter which, after some time, is expected to enter a hydrodynamic regime (except in dS \cite{Casalderrey-Solana:2020vls}, see below).
For $\Lambda=0$ this is illustrated in Fig.~\ref{fig:pumukel}, which shows the evolution of the pressure/energy ratio for three different initial conditions, IC1, IC2 and IC3.
For IC2 and IC3 we added respectively $+2$ and $-1.5$ to the subtracted $\phi(r,0)$ of IC1.
This leads to evolutions that are just about numerically stable and hence as far from equilibrium as our code allows.
The blue curves are the holographic results.
The difference with the viscous hydrodynamic approximation \cite{Casalderrey-Solana:2020vls} (green curves) at early times shows that the initial dynamics is far from equilibrium. 
After $\Delta t \approx 2$ the evolution becomes well described by viscous hydrodynamics, consistently with a hydrodynamization time of $\mathcal{O}(1/T)$ \cite{Heller:2011ju,Heller:2012km}.
The comparison to ideal hydrodynamics in the right panel of Fig.~\ref{fig:pumukel} shows that viscous corrections can be sizable even at late times.

\begin{figure}
	\begin{center}
    \includegraphics[height=0.5\textwidth]{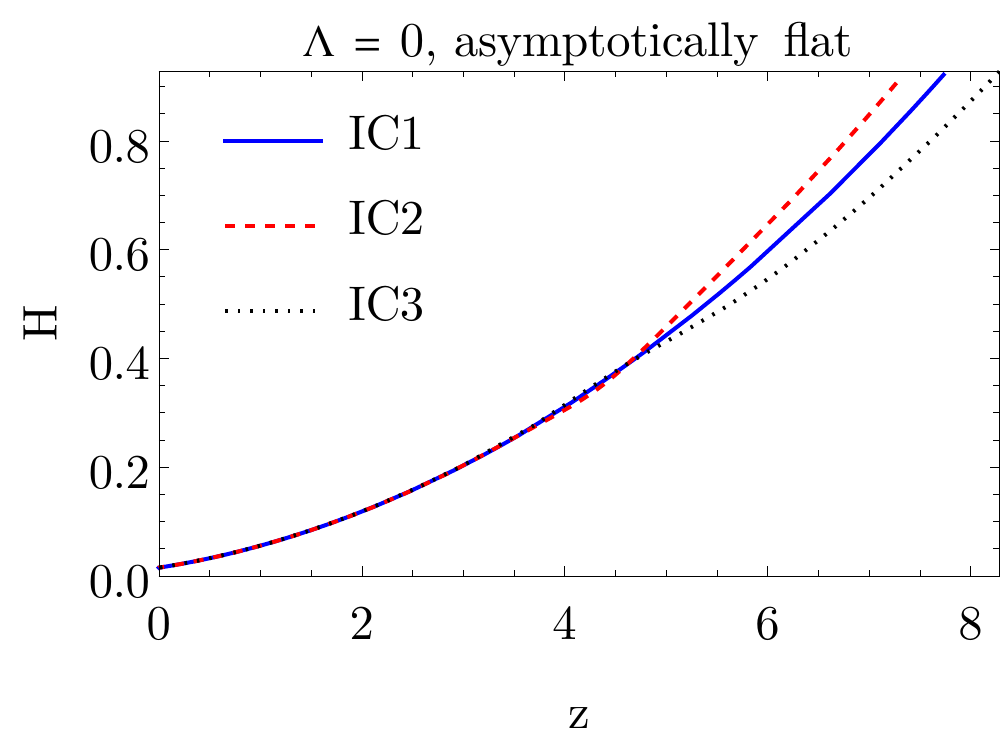}
    \caption{\small 
    Evolution of the Hubble rate as function of redshift $z$ for the three different initial conditions presented in Fig.~\ref{fig:pumukel}.
    }
    \label{fig:zHT}
        	\end{center}
\end{figure}

The initial far-from-equilibrium period leaves an imprint on the scale factor.
This is illustrated in Fig.~\ref{fig:zHT}, which shows the Hubble rate for the three evolutions of Fig.~\ref{fig:pumukel} as a function of the redshift \mbox{$z(t)=a(t_{\rm{obs}})/a(t)-1$}.
The time $t_{\rm obs}$ is defined for each curve by the physical condition that $\mathcal{E}$ reaches some late-time value, in this case $\mathcal{E}(t_{\rm obs})=0.02$.
At small redshift the evolutions are equivalent as a consequence of the applicability of hydrodynamics at late times shown in Fig.~\ref{fig:pumukel}.
In contrast, at large redshift the far-from-equilibrium dynamics at early times leads to significantly different Hubble rates.

\begin{figure}
	\begin{center}
    \includegraphics[height=0.5\textwidth]{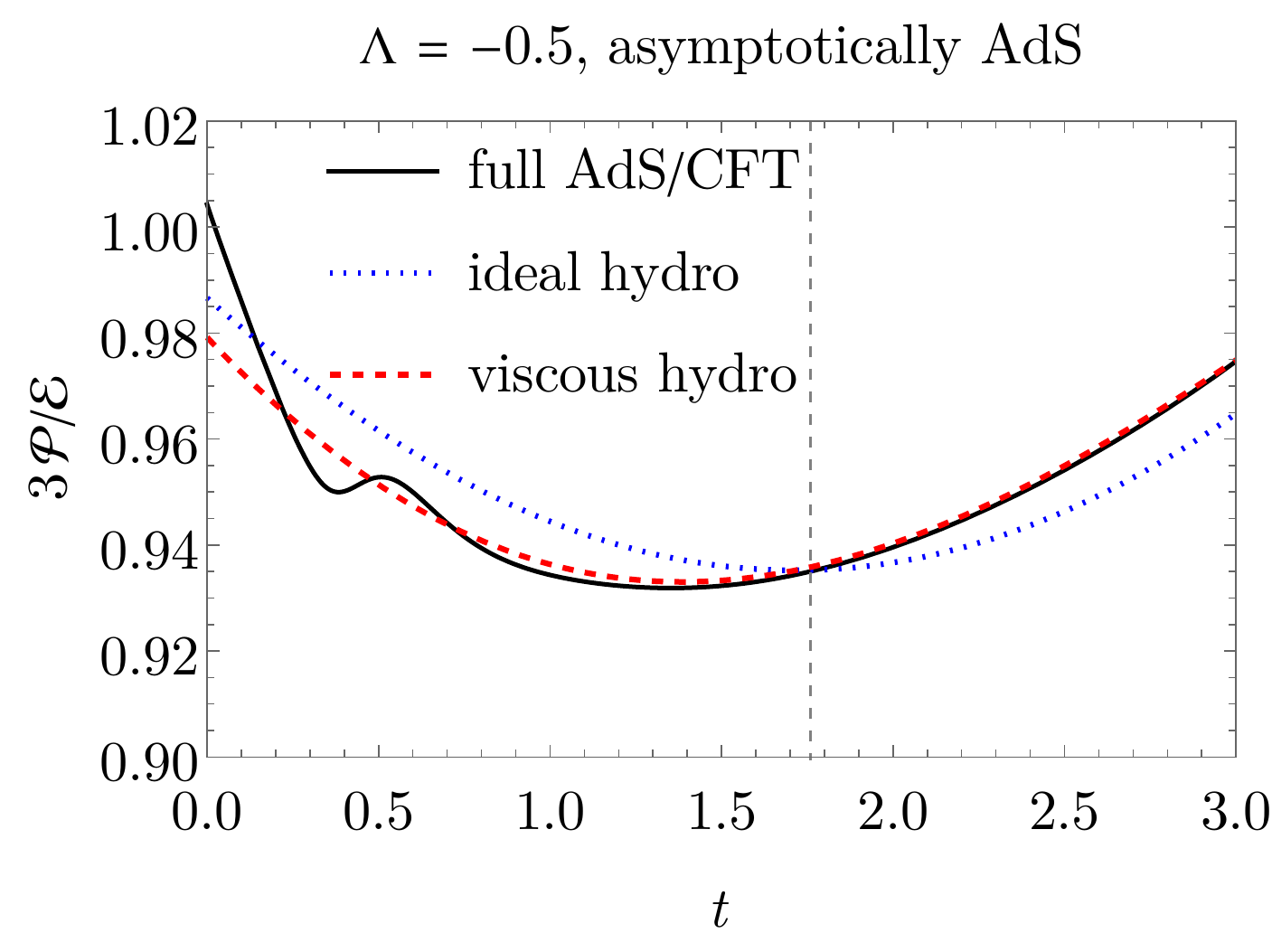}
    \caption{\small 
    Comparison between the holographic result for the pressure/energy ratio (black) and the ideal (blue) and viscous (red)  approximations, for $G=1/2500$ and $\Lambda=-0.5$.
    }
    \label{fig:Lambdam50}
    	\end{center}
\end{figure}

In Fig.~\ref{fig:Lambdam50} we show the analogous results for \mbox{$\Lambda=-0.5$}.
The dashed, grey line marks the time where $\mathcal{E}$ reaches a minimum and $H=0$. 
The entire evolution is well described by viscous hydrodynamics. As above, viscous corrections are non-negligible at late times.

\begin{figure}
	\begin{center}
    \includegraphics[height=0.5\textwidth]{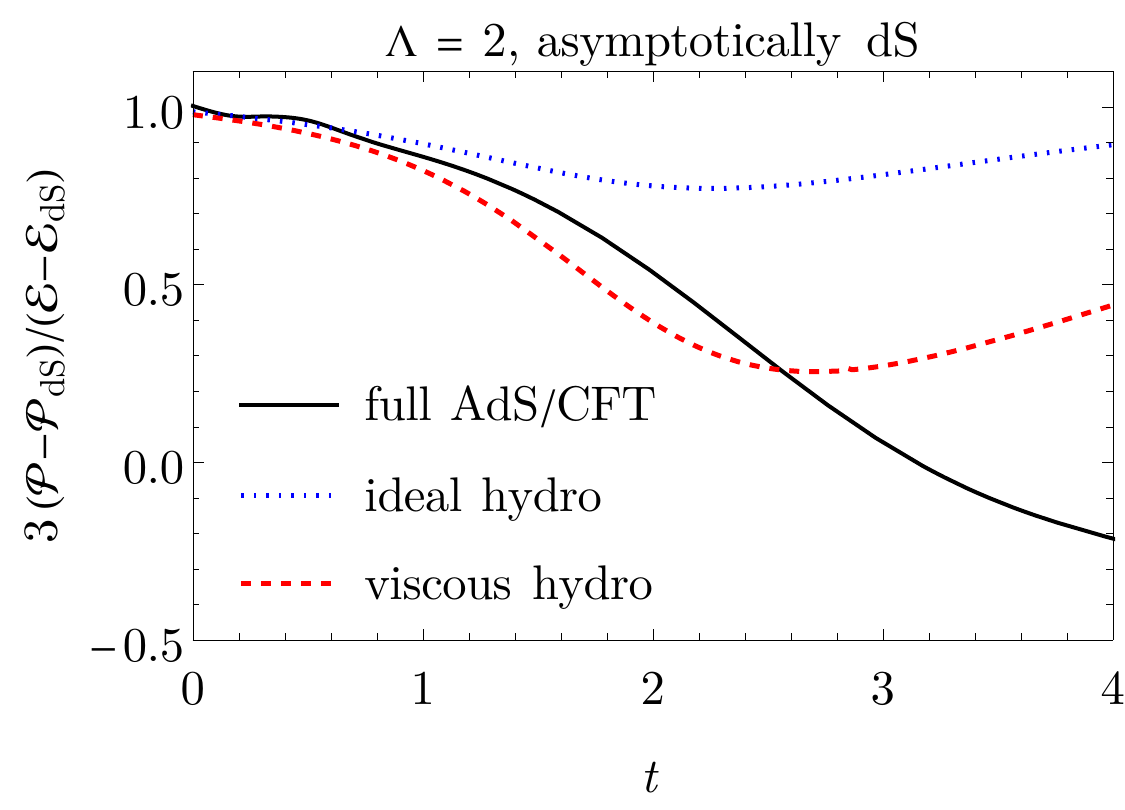}
    \caption{\small Same as in Fig.~\ref{fig:Lambdam50} but for $\Lambda=2$.}
    \label{fig:Lambda200}
    		\end{center}
\end{figure}

Fig.~\ref{fig:Lambda200} illustrates the asymtotically dS case.
At late times the backreaction is dominated by the cosmological constant, which here includes a Casimir contribution that we subtract in the plot.
Once the expansion has diluted the energy density so that $\mathcal{E}-\mathcal{E}_{\rm dS} \lesssim H^4$, the system is driven out of equilibrium and the hydrodynamic approximation ceases to be valid, as expected from the non-backreacted  analysis \cite{Casalderrey-Solana:2020vls}.

%%%%%%%%%%%%%%%%%%%%%%%%%%%
\section{Discussion}
\label{disc}
%%%%%%%%%%%%%%%%%%%%%%%%%%%%
We have provided the first example of holographic time evolution with dynamical boundary gravity in which both the bulk and the boundary geometries are constructed dynamically, one time step at a time. 

In order to illustrate our approach in the simplest possible setting, we have focused on homogeneous and isotropic states.  However, we expect that our scheme can be generalised to situations with no symmetry assumptions. Our work thus suggests new  possible applications of holography that we will develop elsewhere.  Here we just close with brief comments on two of them.

Inflation could be studied by promoting the boundary value of the bulk scalar field to a dynamical boundary scalar field which would play the role of the inflaton.
This would allow us to use holography to study e.g.~the pre- and re-heating processes at the end of inflation \cite{Kofman:1994rk,Kofman:1997yn}.

In the absence of symmetry assumptions, cosmological backgrounds are expected to suffer from instabilities \cite{Mukhanov:1996ak}.
This has been  studied holographically in the linear approximation \cite{Chesler:2020exl}.
Our approach would allow us to determine the endpoint of these instabilities deep into the nonlinear regime.

\acknowledgments
It is a pleasure to thank R.~Emparan, J.~Garriga, G.~Horowitz, E.~Kiritsis and J.~Mas for discussions.  
JCS and DM are supported by grants SGR-2017-754, PID2019-105614GB-C21, PID2019-105614GB-C22 and the ``Unit of Excellence MdM 2020-2023'' award to the Institute of Cosmos Sciences (CEX2019-000918-M).

%%%%%%%%%%%%%%%%%%%%%%%%%%%%%%%%%%%%%%%%%%
\bibliography{main5}{}
\bibliographystyle{JHEP}
\end{document}